\documentclass[conference]{IEEEtran}

\usepackage[utf8]{inputenc} 
\usepackage[T1]{fontenc}    
\usepackage{url}            
\usepackage{booktabs}       
\usepackage{nicefrac}       
\usepackage{microtype}      
\usepackage{xcolor}         
\usepackage{wrapfig}
\usepackage{graphicx}

\usepackage{amsmath, amssymb, amsfonts}
\usepackage{float}
\usepackage{framed}
\usepackage{enumerate}
\usepackage{algorithm}
\usepackage{algpseudocode}
\usepackage{adjustbox}
\usepackage{textcomp}
\usepackage{dsfont}  

\usepackage{multirow}
\usepackage{array}
\usepackage{tabularx}
\usepackage{nameref}
\usepackage{makecell}
\usepackage{cancel}
\usepackage{fancyhdr}
\usepackage{dirtytalk}
\usepackage{cite}
\def\BibTeX{{\rm B\kern-.05em{\sc i\kern-.025em b}\kern-.08em
    T\kern-.1667em\lower.7ex\hbox{E}\kern-.125emX}}

\ifCLASSOPTIONcompsoc 
  \usepackage[caption=false,font=normalsize,labelfont=sf,textfont=sf]{subfig}
\else
  \usepackage[caption=false,font=footnotesize]{subfig}
\fi

\newcommand{\negspace}{\vspace{-0.50\baselineskip}}

\title{Exploring QUIC Dynamics: A Large‑Scale Dataset for Encrypted Traffic Analysis}

\author{\IEEEauthorblockN{Barak Gahtan\IEEEauthorrefmark{1}\thanks{\IEEEauthorrefmark{1}Both authors contributed equally.}}
\IEEEauthorblockA{\textit{Technion, CS dept, Israel} \\
barakgahtan@cs.technion.ac.il}
\and
\IEEEauthorblockN{Robert J.~Shahla\IEEEauthorrefmark{1}}
\IEEEauthorblockA{\textit{Technion, CS dept, Israel} \\
shahlarobert@cs.technion.ac.il}
\and
\IEEEauthorblockN{Reuven Cohen}
\IEEEauthorblockA{\textit{Technion, CS dept, Israel} \\
rcohen@cs.technion.ac.il}
\and
\IEEEauthorblockN{Alex M.~Bronstein}
\IEEEauthorblockA{\textit{Technion, CS dept, Israel} \\
bron@cs.technion.ac.il}
}
\IEEEoverridecommandlockouts

\begin{document}
\maketitle
\begin{abstract}
The increasing adoption of the QUIC transport protocol has transformed encrypted web traffic, necessitating new methodologies for network analysis. However, existing datasets lack the scope, metadata, and decryption capabilities required for robust benchmarking in encrypted traffic research.

We introduce VisQUIC, a large-scale dataset of 100,000 labeled QUIC traces from over 44,000 websites, collected over four months. Unlike prior datasets, VisQUIC provides SSL keys for controlled decryption, supports multiple QUIC implementations, and introduces a novel image-based representation that enables machine learning (ML)-driven encrypted traffic analysis, along with standardized benchmarking tools, ensuring reproducibility.

To demonstrate VisQUIC’s utility, we present a benchmarking task for estimating HTTP/3 responses in encrypted QUIC traffic, achieving 97\% accuracy using only observable packet features. By publicly releasing VisQUIC, we provide an open foundation for advancing encrypted traffic analysis, QUIC security research, and network monitoring.
\end{abstract}

\begin{IEEEkeywords}
QUIC, HTTP/3, Encrypted Traffic Analysis, Machine Learning, Deep Learning, Network Security, Benchmarking, Traffic Monitoring.
\end{IEEEkeywords}

\section{Introduction}\label{intro}
The widespread adoption of Quick UDP Internet Connections (QUIC) by major platforms such as Google, Facebook, and Cloudflare has transformed web traffic, improving both security and performance~\cite{8999454,Yu2021Dissecting,Serreli2023}. Unlike TCP, QUIC integrates encryption at the transport layer~\cite{rfc9000}, enhancing security but complicating network analysis. Traditional traffic monitoring methods relying on unencrypted headers and payload inspection are now ineffective, necessitating novel approaches for analyzing encrypted traffic~\cite{geiginger2021classification,zhang2023application}.

Despite QUIC’s widespread adoption, large-scale datasets capturing its encrypted nature remain scarce~\cite{Luxemburk2023}. Existing datasets often lack metadata, fail to represent QUIC’s diverse implementations, or omit structured benchmarking tools, limiting their usefulness for ML-driven encrypted traffic analysis.

To bridge this gap, we introduce VisQUIC, a dataset designed for encrypted traffic analysis and benchmarking. VisQUIC includes 100,000+ labeled QUIC traces from 44,000+ websites, collected over four months, with SSL keys enabling controlled decryption for research. By spanning diverse network conditions, VisQUIC supports comprehensive studies on QUIC security, traffic classification, and performance optimization.

Key Contributions of VisQUIC: \begin{itemize} \item \textbf{Diverse QUIC Coverage:} Multiple implementations across varied network environments. \item \textbf{Controlled Decryption:} SSL keys for encrypted traffic analysis. \item \textbf{Image-Based Representation:} Structured visual formats for ML applications. \item \textbf{Standardized Benchmarking:} Tools and metrics for reproducible evaluation. \end{itemize}

VisQUIC enables data-driven research with a novel image-based representation of QUIC traffic, allowing ML models to identify traffic patterns without full decryption.

To illustrate its utility, we introduce a benchmark algorithm for estimating HTTP/3 response counts in encrypted QUIC connections. Leveraging VisQUIC’s image-based transformation, this algorithm achieves 97\% accuracy, demonstrating the dataset’s benchmarking potential. However, this paper primarily focuses on presenting VisQUIC, with benchmarking serving as an illustrative application. VisQUIC is publicly available via our GitHub repository\footnote{\url{https://github.com/robshahla/VisQUIC}}, ensuring accessibility and reproducibility. It includes detailed documentation and standardized evaluation tools to support research in network security, encrypted traffic analysis, and performance modeling.

Beyond HTTP/3 response estimation, VisQUIC facilitates standardized evaluations for protocol classification, encrypted traffic fingerprinting, congestion control analysis, and anomaly detection. By establishing a reproducible dataset with structured benchmarks, VisQUIC advances research in network security, privacy-preserving ML, and encrypted traffic modeling.

This paper is structured as follows: Section \ref{relatedwork} reviews related datasets and highlights VisQUIC’s unique contributions. Section \ref{solution} details the dataset collection methodology. Section \ref{benchmark} presents the benchmark algorithm as an illustrative use case. Finally, Section \ref{conclusion} discusses implications and future work.
\section{Related Work}\label{relatedwork}
As QUIC adoption grows, research on its traffic analysis has expanded significantly~\cite{almuhammadi2023quic}. Yet, \textbf{progress remains limited due to the lack of publicly available datasets} with both encrypted QUIC traces and structured metadata for benchmarking ML models~\cite{geiginger2021classification,zhang2023application}. Existing datasets lack metadata, HTTP/3 coverage, or decryption support, reducing effectiveness for systematic benchmarking.

\begin{figure}[t] 
\centering 
\includegraphics[width=1\linewidth]{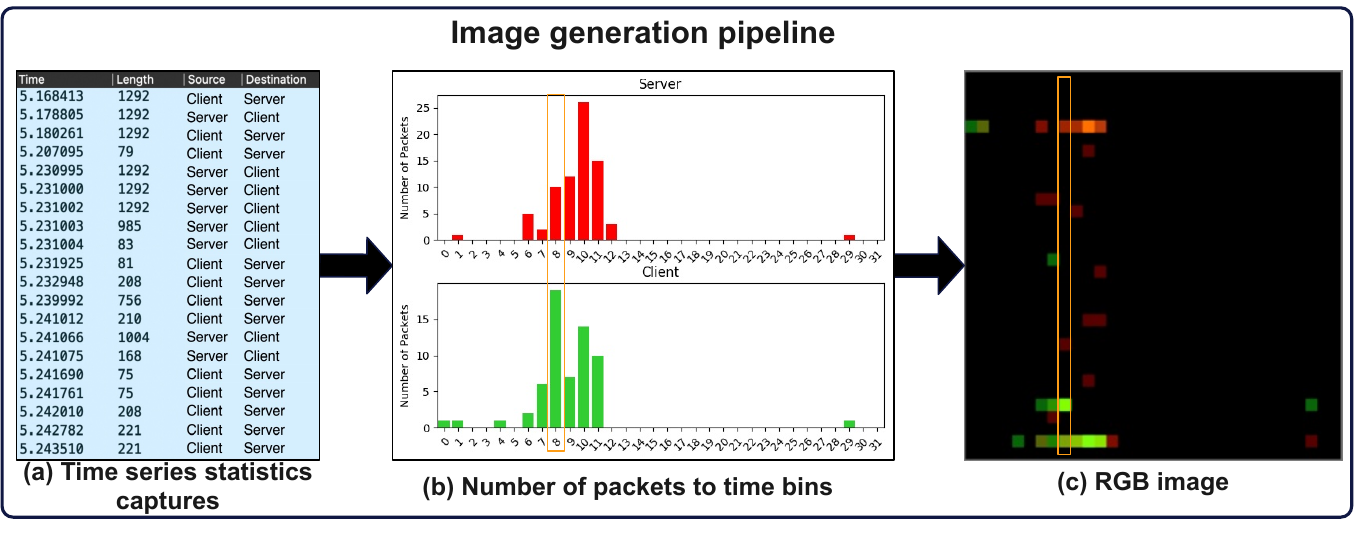} 
\caption{Constructing an image from a QUIC trace. (a) Raw packet metadata captures timing, size, and direction. (b) Packets are binned by time and length, creating histograms for client-to-server (green) and server-to-client (red) traffic. (c) The final RGB representation preserves temporal relationships and directionality, where pixel intensity indicates packet density.}
\negspace
\negspace
\label{fig:image-const} 
\end{figure}

CESNET-QUIC22~\cite{luxemburk2023cesnet} captures 153 million QUIC connections from an ISP but limits metadata to only the first 30 packets and lacks HTTP/3 data and SSL keys, preventing full QUIC session reconstruction. In~\cite{smith2021website} TCP/QUIC traces from VPN gateways were introduced, but reliance on VPN traffic introduces inconsistencies in latency, congestion control, and routing, reducing benchmark reliability. Similarly, CAIDA’s dataset~\cite{caida_passive_dataset} provides backbone traffic traces but lacks QUIC payloads, limiting studies on encryption, multiplexing, and application-layer behaviors.

Beyond web browsing, QUIC is widely used in mobile applications, WebRTC, and cloud services~\cite{mosaic, mvfst, chromiumquiche, quiche}. Yet, existing datasets often lack this diversity, focusing on a single QUIC implementation or data collection environment.

While previous datasets offer insights into QUIC traffic, they fail to serve as dedicated benchmarks for encrypted traffic classification, QoE-aware analysis, and next-generation network performance evaluation~\cite{zhang2023application}. An effective benchmark must provide comprehensive encrypted traffic samples with metadata, reproducible model evaluation~\cite{Shen2023}, and privacy-preserving accessibility while maintaining security standards. Recent efforts, such as NetBench, highlight the need for structured approaches to benchmarking encrypted traffic analysis~\cite{Qian2024}.

To address these gaps, we introduce \textbf{VisQUIC}, a dataset explicitly designed for large-scale encrypted traffic benchmarking. VisQUIC surpasses prior datasets by offering \textbf{100,000+ labeled QUIC traces from 44,000+ websites} collected over four months. It includes SSL keys for controlled decryption, diverse QUIC implementations, and a novel image-based transformation that enables ML applications~\cite{Ding2016A}. Unlike prior datasets that offer limited metadata or partial packet captures, VisQUIC provides structured QUIC traces, allowing researchers to analyze encrypted traffic across multiple QUIC implementations in a controlled yet realistic setting.

\section{VisQUIC: A Dataset for Encrypted QUIC Traffic Analysis}\label{solution}
\subsection{Dataset Collection and Structure}
To ensure broad coverage, we collected QUIC traces from two residential networks across different continents, capturing diverse network conditions and geographical variations. The data collection process spanned all hours of the day, allowing performance benchmarking under varying congestion, routing, and network scenarios.

\begin{figure}[t!]
    \centering
    \subfloat[$16 \times 16$\label{fig:161}]{
        \includegraphics[width=0.15\textwidth]{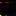}
    }
    \subfloat[$32 \times 32$\label{fig:321}]{
        \includegraphics[width=0.15\textwidth]{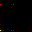}
    }
    \subfloat[$64 \times 64$\label{fig:641}]{
        \includegraphics[width=0.15\textwidth]{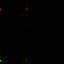}
    }\hfill
    \subfloat[$128 \times 128$\label{fig:1281}]{
        \includegraphics[width=0.23\textwidth]{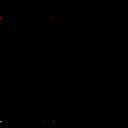}
    }\hfill
    \subfloat[$256 \times 256$\label{fig:2561}]{
        \includegraphics[width=0.23\textwidth]{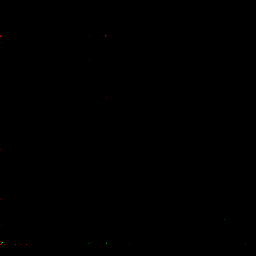}
    }
    \caption{Comparison of QUIC image representations at different resolutions. Lower resolutions lose fine packet detail, while higher resolutions capture intricate temporal patterns.}
    \label{fig:different-responses1}
    \negspace
\end{figure}

We built the dataset by probing HTTP/3-enabled websites from the Tranco list~\cite{tranco,Pochat2019}. Each website was accessed using Headless Chrome~\cite{headless} in incognito mode with caching disabled to ensure consistency. To eliminate session resumption artifacts, websites were accessed sequentially, ensuring independent QUIC connections. This approach provides a clean, reproducible dataset representing real-world web interactions.

Unlike video streaming datasets driven by adaptive bitrate algorithms, VisQUIC targets web browsing traffic with dynamic content loading, third-party services, and server-driven responses, requiring diverse packet structures. While the current dataset is Chrome-based, future work could extend it to include additional browsers to capture implementation variations.

QUIC traffic was captured with \texttt{Tshark}~\cite{merino2013instant} in PCAP format, retaining only QUIC packets for encrypted traffic analysis. Each PCAP file is paired with its SSL keys for controlled decryption. This feature allows encrypted traffic classification while maintaining interpretability~\cite{Jo2024Encrypted,Wilkens2021Passive}.

\subsection{Image-Based Representation for ML Applications}
Beyond raw traces, VisQUIC introduces an image-based representation designed to support ML applications. This transformation builds upon prior work in network traffic visualization ~\cite{shapira2019flowpic, golubev2022image,Swathi2022}, offering a structured way to analyze QUIC flows without requiring full decryption~\cite{Yu2011}. Deep learning approaches have demonstrated the effectiveness of network traffic image representations for security applications, such as anomaly detection and malware classification~\cite{Wang2018}. Moreover, recent advancements in bidirectional flow-based image representations further refine network traffic categorization, enabling high-accuracy encrypted traffic classification without exposing sensitive payload data~\cite{Jiang2024}. These developments highlight the increasing importance of image-based traffic representations in modern network analysis frameworks~\cite{Marwah2022}.

\begin{table}
\caption{Summary statistics of QUIC traces and the number of images per dataset for each web server.}
\begin{tabular}{|p{2.2cm}|p{1.3cm}|p{0.9cm}|p{1.2cm}|c|}
\hline
\centering \textbf{Web Server} & \centering \textbf{Websites} & \centering \textbf{Traces} & \centering \textbf{$T=0.1$} & \textbf{$T=0.3$} \\ \hline
\centering youtube & \centering 399 & \centering 2,109 & \centering 139,889 & 54,659 \\ \hline
\centering semrush & \centering 1,785 & \centering 9,489 & \centering 474,716 & 221,477 \\ \hline
\centering discord & \centering 527 & \centering 7,271 & \centering 623,823 & 235,248 \\ \hline
\centering instagram & \centering 3 & \centering 207 & \centering 17,003 & 7,112 \\ \hline
\centering mercedes-benz & \centering 46 & \centering 66 & \centering 9,987 & 2,740 \\ \hline
\centering bleacherreport & \centering 1,798 & \centering 8,497 & \centering 781,915 & 331,530 \\ \hline
\centering nicelocal & \centering 1,744 & \centering 1,666 & \centering 148,254 & 48,900 \\ \hline
\centering facebook & \centering 13 & \centering 672 & \centering 25,919 & 10,988 \\ \hline
\centering pcmag & \centering 5,592 & \centering 13,921 & \centering 1,183,717 & 385,797 \\ \hline
\centering logitech & \centering 177 & \centering 728 & \centering 56,792 & 28,580 \\ \hline
\centering google & \centering 1,341 & \centering 2,149 & \centering 81,293 & 29,068 \\ \hline
\centering cdnetworks & \centering 902 & \centering 2,275 & \centering 207,604 & 85,707 \\ \hline
\centering independent & \centering 3,340 & \centering 3,453 & \centering 176,768 & 68,480 \\ \hline
\centering cloudflare & \centering 26,738 & \centering 44,700 & \centering 1,347,766 & 341,488 \\ \hline
\centering jetbrains & \centering 35 & \centering 1,096 & \centering 34,934 & 18,470 \\ \hline
\centering pinterest & \centering 43 & \centering 238 & \centering 6,465 & 2,360 \\ \hline
\centering wiggle & \centering 4 & \centering 0 & \centering 0 & 0 \\ \hline
\centering cnn & \centering 27 & \centering 2,127 & \centering 91,321 & 59,671 \\ \hline
\end{tabular}
\label{tab:combined-table1}
\end{table}
\begin{figure*}[!t]
    \centering
    \includegraphics[width=0.8\textwidth]{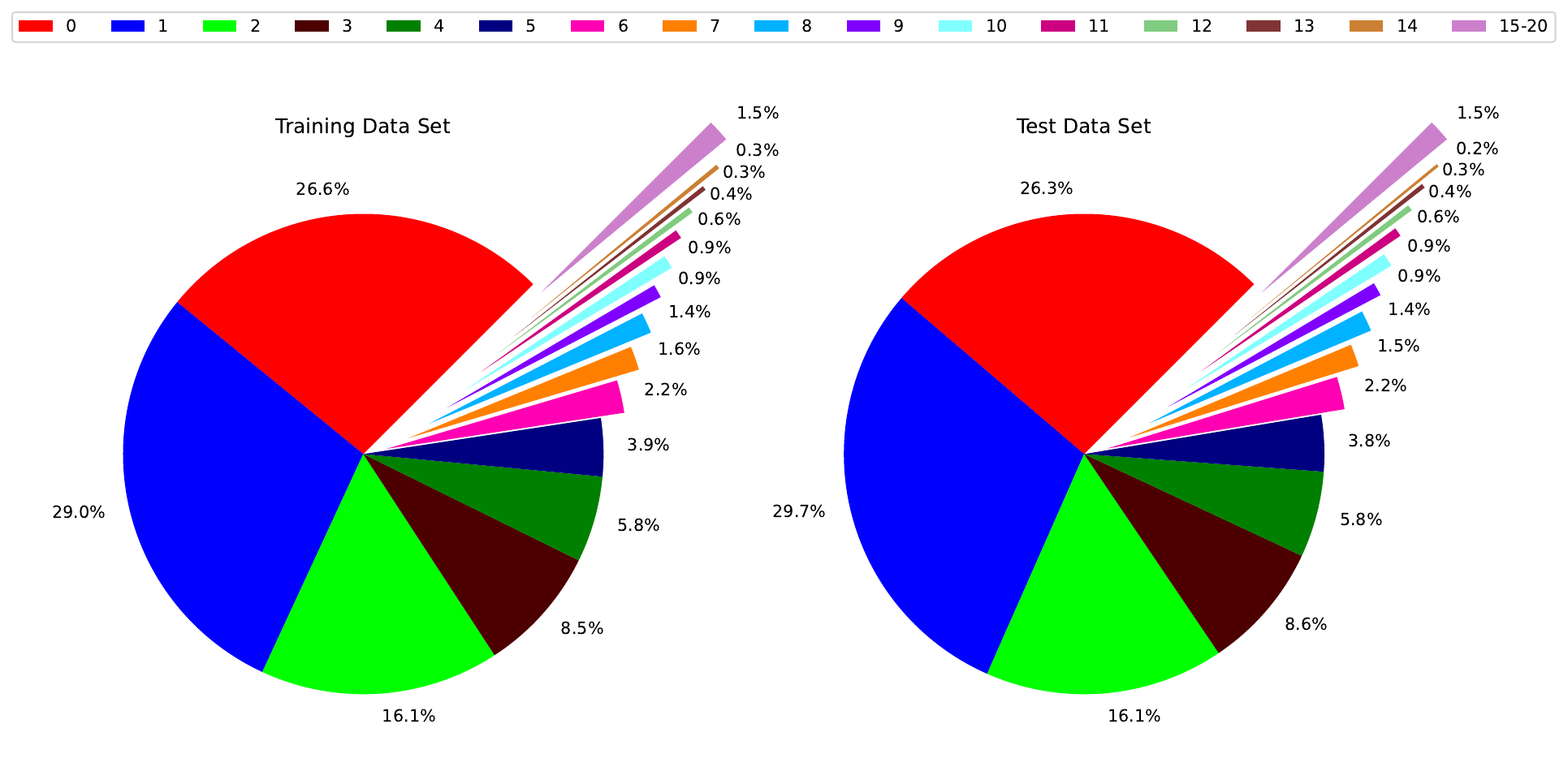}
    \caption{Response distribution for training and evaluation datasets with a $T=0.1$-second sliding window.}
    \label{fig:response-distribution-dataset}
    \negspace
    \negspace
\end{figure*}


Fig.~\ref{fig:image-const} shows how QUIC traces are converted into images. Key metadata such as arrival time, packet size, and direction (client-to-server or server-to-client) is extracted from each packet and organized into structured histograms. The data is binned along two axes—time and packet size—capturing both the temporal and volumetric characteristics of the traffic in a grid. Packets transmitted in different directions are mapped to separate color channels: red for server-to-client and green for client-to-server. This multi-channel encoding improves the flow direction differentiation and multiplexing in HTTP/3 traffic compared to prior single-channel grayscale methods. The effectiveness of image representations in QUIC traffic analysis depends on three key factors:

\textbf{Window Length ($T$).} Defines each image's temporal span, balancing detail and computational cost. Short windows capture fine-grained packet details but increase storage and processing demands. Longer windows aggregate traffic over time, reducing image count but potentially hiding transient behaviors.

\textbf{Image Resolution.} Determines the level of structural detail captured. Higher resolutions enhance fine-grained feature extraction but require greater computational resources. The optimal resolution depends on the trade-off between accuracy and efficiency for different ML models.

\begin{table}[t!]
\centering
\caption{CAP results for known web servers, using five random training/test splits at \( T=0.1 \) and \( T=0.3 \).}
\begin{tabular}{|c|c|c|c|c|}
\hline
\textbf{Iteration} & \multicolumn{2}{c|}{\textbf{\( T=0.1 \)}} & \multicolumn{2}{c|}{\textbf{\( T=0.3 \)}} \\ \hline
 & \textbf{\( \pm1 \)} & \textbf{\( \pm2 \)} & \textbf{\( \pm1 \)} & \textbf{\( \pm2 \)} \\ \hline
1 & 0.93 & 0.97 & 0.91 & 0.96 \\ \hline
2 & 0.92 & 0.96 & 0.90 & 0.97 \\ \hline
3 & 0.93 & 0.98 & 0.91 & 0.95 \\ \hline
4 & 0.94 & 0.97 & 0.92 & 0.93 \\ \hline
5 & 0.91 & 0.96 & 0.92 & 0.94 \\ \hline
\end{tabular}
\label{tab:CAP_results_known}
\negspace
\end{table}

\textbf{Normalization Strategy.} Affects interpretability and model performance. Per-window normalization highlights short-term variations, making it effective for detecting rapid traffic fluctuations, anomalies, and congestion patterns. Per-trace normalization captures long-term trends but may obscure local deviations, making it more suitable for fingerprinting and congestion control analysis. The choice depends on the target application and computational constraints.

Fig.~\ref{fig:different-responses1} shows the impact of resolution on image representation. At lower resolutions (Fig.~\ref{fig:different-responses1}\subref{fig:161}), packet aggregation across both directions (red and green channels) results in a yellow pixel. Higher resolutions (Fig.~\ref{fig:different-responses1}\subref{fig:321}) and~\ref{fig:different-responses1}\subref{fig:641}) improve directional flow differentiation, enhancing interpret-ability for ML applications.

Lower resolutions ($16\times16$, $32\times32$) are efficient for real-time classification, while higher ones ($128\times128$, $256\times256$) preserve intricate packet interactions, benefiting fine-grained anomaly detection, encrypted traffic fingerprinting, and HTTP/3 multiplexing analysis. Researchers must balance computational cost with the level of structural detail required for their task.

Transforming QUIC traffic into image representations provides a structured abstraction, enabling ML models to recognize traffic patterns without decryption. Prior works such as FlowPic~\cite{shapira2019flowpic} and~\cite{golubev2022image} show that network traffic image encoding enhances classification and anomaly detection. VisQUIC extends these approaches to QUIC and HTTP/3 by offering a multi-channel representation capturing packet timing, direction, and volumetric flow for deep learning applications.

Unlike traditional feature-based or packet-level analysis, image-based representations remove the need for manual feature engineering. Deep learning models can infer traffic patterns directly from images, making this approach well-suited for tasks like encrypted traffic classification, anomaly detection, and congestion analysis. VisQUIC’s structured format also enables transfer learning across QUIC implementations, allowing models trained on one (e.g., Chromium QUIC~\cite{chromiumquiche}) to generalize to others (e.g., mvfast~\cite{mvfst}).


\subsection{Dataset Scope and Accessibility}
VisQUIC captures QUIC traffic from multiple implementations, including Chromium QUIC~\cite{chromiumquiche}, mvfst~\cite{mvfst}, and quiche~\cite{quiche}. The dataset encompasses a diverse range of services, from social media platforms to content delivery networks and independent publishers, ensuring a representative sample of modern QUIC usage. Table~\ref{tab:combined-table1} provides an overview of the collected traces across different web services, highlighting the dataset's breadth and diversity.

\subsection{Potential Applications of VisQUIC}
VisQUIC is a valuable resource for both networking and ML communities, enabling real-world analysis of QUIC and HTTP/3 traffic. By providing structured metadata, encrypted traces, and an image-based representation, it supports applications in 5G/6G network security, QoS-aware encrypted traffic classification, and congestion control.

From a networking view, VisQUIC supports research on anomaly detection, DDoS mitigation, and congestion control. It enables precise round-trip time estimation, helps ISPs identify encrypted traffic patterns, optimize resource allocation~\cite{shahla2024trafficgrinder}, and assess QUIC’s impact on intrusion detection systems.

For ML, VisQUIC offers structured image representations for classification and regression. Researchers can study image resolution impact on deep learning models, use transfer learning across QUIC implementations (e.g., Chromium~\cite{chromiumquiche} vs. quiche~\cite{quiche}), and develop CNNs and Vision Transformers for encrypted traffic fingerprinting.

Future directions include expanding VisQUIC to mobile and IoT traffic, developing hybrid privacy-preserving frameworks, and introducing new benchmarking tasks beyond HTTP/3 response estimation. VisQUIC provides a reproducible benchmarking foundation, promoting standardized evaluation metrics for encrypted traffic analysis in academia and industry.


\section{Benchmarking HTTP/3 Response Estimation}\label{benchmark}
To showcase VisQUIC’s value as a benchmarking dataset, we present an example task: estimating the number of HTTP/3 responses within encrypted QUIC connections. This demonstrates the feasibility of analyzing encrypted traffic based solely on observable packet characteristics, without plaintext inspection.

Estimating responses in encrypted traffic is critical for multiple reasons. First, it assesses a model's ability to extract meaningful patterns from encrypted flows. Second, it has practical applications in load balancing, where identifying heavy flows aids in optimizing server selection~\cite{shahla2024trafficgrinder}. Finally, it provides a standardized metric for comparing traffic analysis methods, highlighting the strengths of various ML models.

We evaluate response estimation using VisQUIC’s image-based representation, splitting each server’s traces randomly in $80$:$20$ ratio for training and testing. Five models were trained on different random splits. Fig.~\ref{fig:response-distribution-dataset} shows the response count distribution in our evaluation sets for the $T=0.1$-second sliding window. Similar distribution observed was for $T=0.3$.

\textbf{Benchmark Implementation and Model Training.}  
QUIC traces were transformed into $(32 \times 32)$ structured images using a sliding window approach. The window length \(T\) defines the temporal resolution, with shorter windows preserving fine-grained interaction details, and longer ones capturing broader patterns. Two configurations were evaluated: \(T=0.1\) seconds and \(T=0.3\) seconds, providing insight into the temporal granularity impact on prediction accuracy. 

To mitigate class imbalance, we designed a custom loss function (Appendix) and selectively applied data augmentation to minority classes (response counts between 10 and 20). Because QUIC image representations capture temporal dependencies, non-order-preserving modifications could hinder feature extraction. Therefore, only minimal noise was introduced using a standard deviation of $\sigma=2.55$ (1\% of pixel value), preserving temporal integrity while improving model robustness~\cite{maharana2022review}. Training was performed with a batch size of 64 using the Adam optimizer~\cite{kingma2014adam} and a ReduceLROnPlateau scheduler, which reduced the learning rate by 30\% upon reaching a validation-loss plateau. Early stopping was applied to prevent overfitting.

While this paper presents an HTTP/3 response estimation as an example application, VisQUIC is not limited to this task. It is designed to support a range of ML-based encrypted traffic research areas, including QUIC connection fingerprinting, congestion prediction, anomaly detection, and flow classification. Future benchmarks could target identifying QUIC server implementations from encrypted traces, estimating connection latency without plaintext headers, or distinguishing between human-driven from automated traffic. By providing a reproducible dataset and standard evaluation metrics, VisQUIC establishes a foundation for broader encrypted traffic benchmarking.

\begin{figure}[t!]
    \centering
    \subfloat[Window length \( T=0.1 \) second\label{fig4:boxplot01mix}]{
        \includegraphics[width=0.48\linewidth]{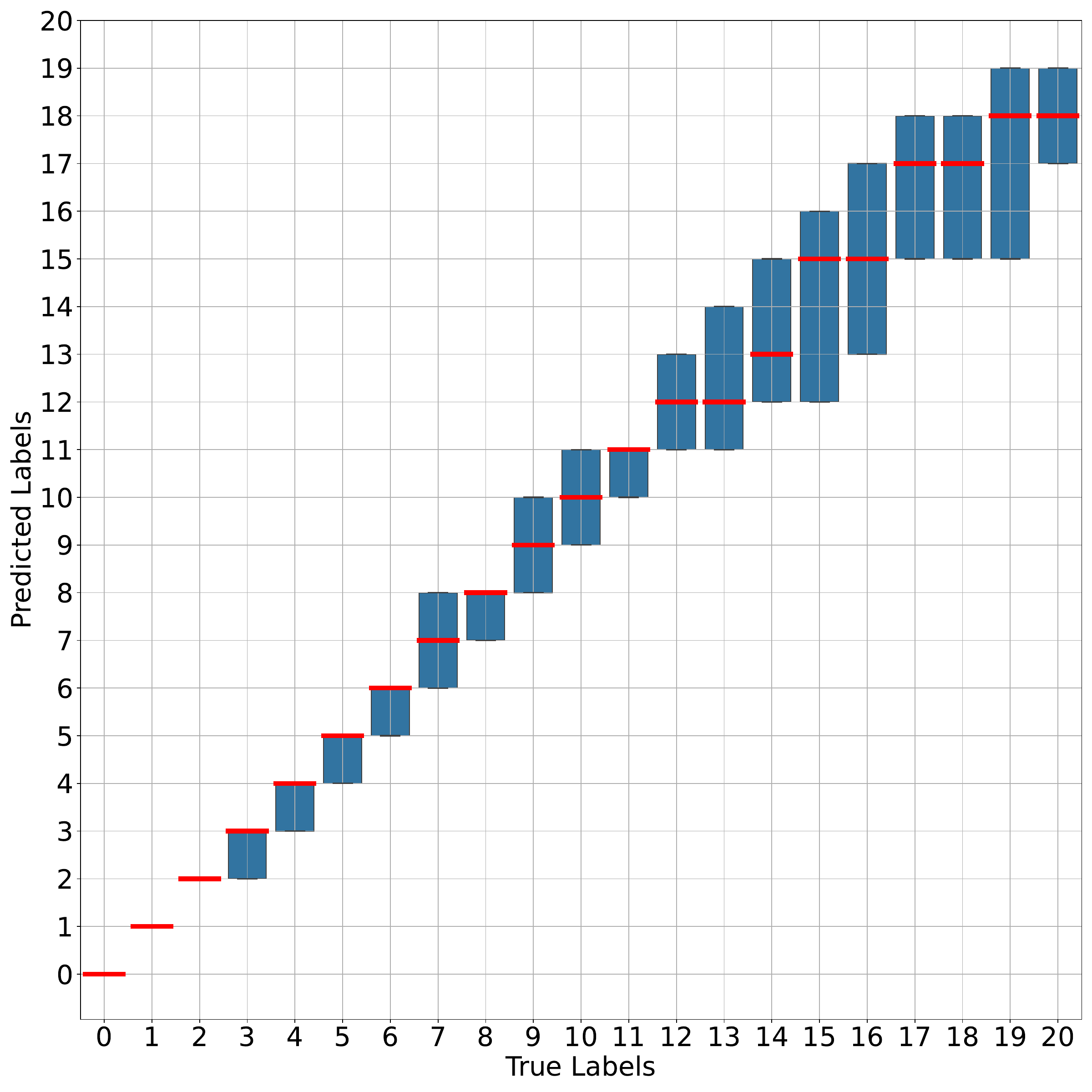}
    }
    \subfloat[Window length \( T=0.3 \) second\label{fig4:boxplot03mix}]{
        \includegraphics[width=0.48\linewidth]{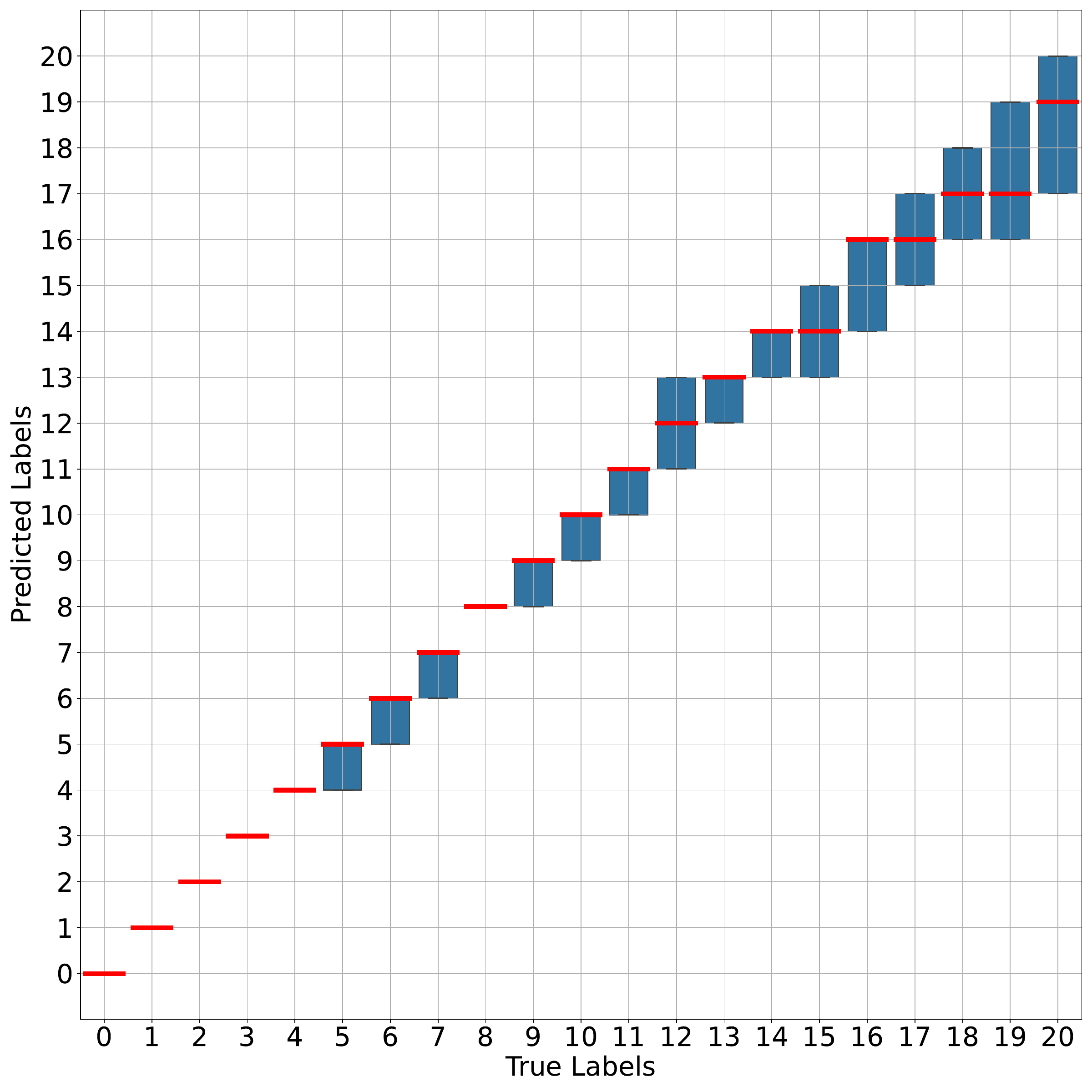}
    }
    \caption{Prediction errors assuming known web servers. Red lines indicate median values; blue boxes represent 25–75\% prediction intervals.}
    \label{fig:box_plot_known}
    \negspace
    \negspace
\end{figure}

\textbf{Evaluation and Results.}  
Fig.~\ref{fig:box_plot_known} presents the distribution of prediction errors across all test traces. At \( T=0.1 \)-second window lengths, lower response counts (0,1,2) exhibit minimal variance, indicating high prediction accuracy for frequent response categories. However, as the true response count increases, the spread of predictions widens due to class imbalance. In contrast, the \( T=0.3 \)-second model achieves stable accuracy up to class 4, with higher response classes maintaining a relatively controlled distribution.


To assess accuracy, we introduce the \textbf{Cumulative Accuracy Profile (CAP)} metric, which quantifies the proportion of predictions within a specified tolerance of the ground truth.
\small
\begin{equation}
  \mathrm{CAP}_{\pm k}(\mathbf{y},\hat{\mathbf{y}})
    = \frac{1}{n}\sum_{i=1}^{n}
      \mathds{1}\bigl(|y_i-\hat{y}_i|\le k\bigr),
\end{equation}
\normalsize
where \( \mathbf{y} \) represents the vector of true class labels, \( \hat{\mathbf{y}} \) denotes model predictions, \( k \) specifies the tolerance level (\(\pm 1\) or \(\pm 2\) classes), and \( n \) is the total number of samples. Unlike exact-match metrics, CAP accounts for near-correct predictions, rewarding those close to the true label.

Table~\ref{tab:CAP_results_known} shows CAP results across five independent training/test splits. The \(T=0.1\) configuration achieves up to 97\% accuracy within a tolerance of \(\pm 2\) responses, while the \(T=0.3\) configuration shows comparable but slightly lower performance.

\begin{figure}[t]
    \centering
    \subfloat[Window length $T=0.1$-sec\label{fig4:scatter_trace01}]{
        \includegraphics[width=0.49\linewidth]{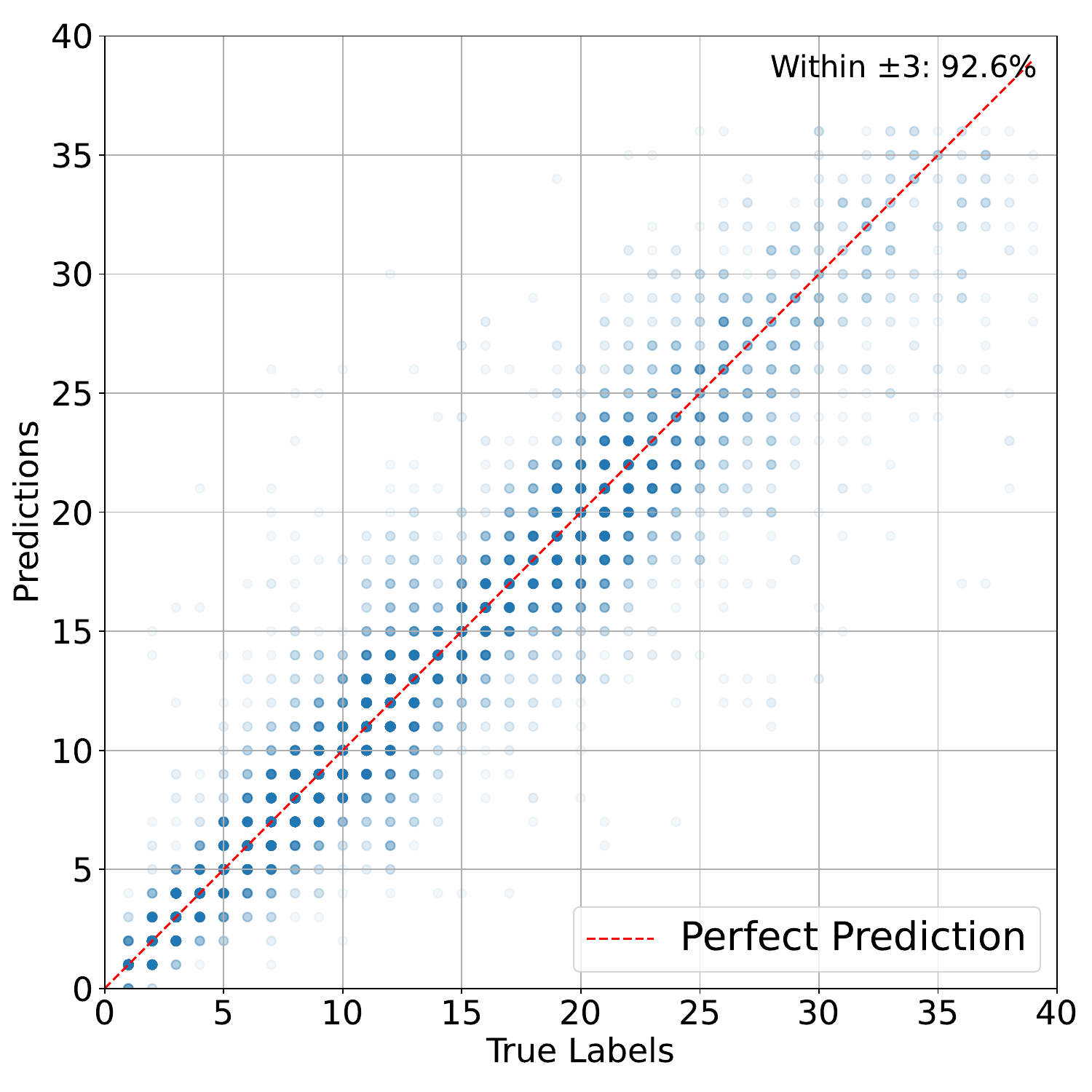}
    }
    \subfloat[Window length $T=0.3$-sec\label{fig4:scatter_trace03}]{
        \includegraphics[width=0.49\linewidth]{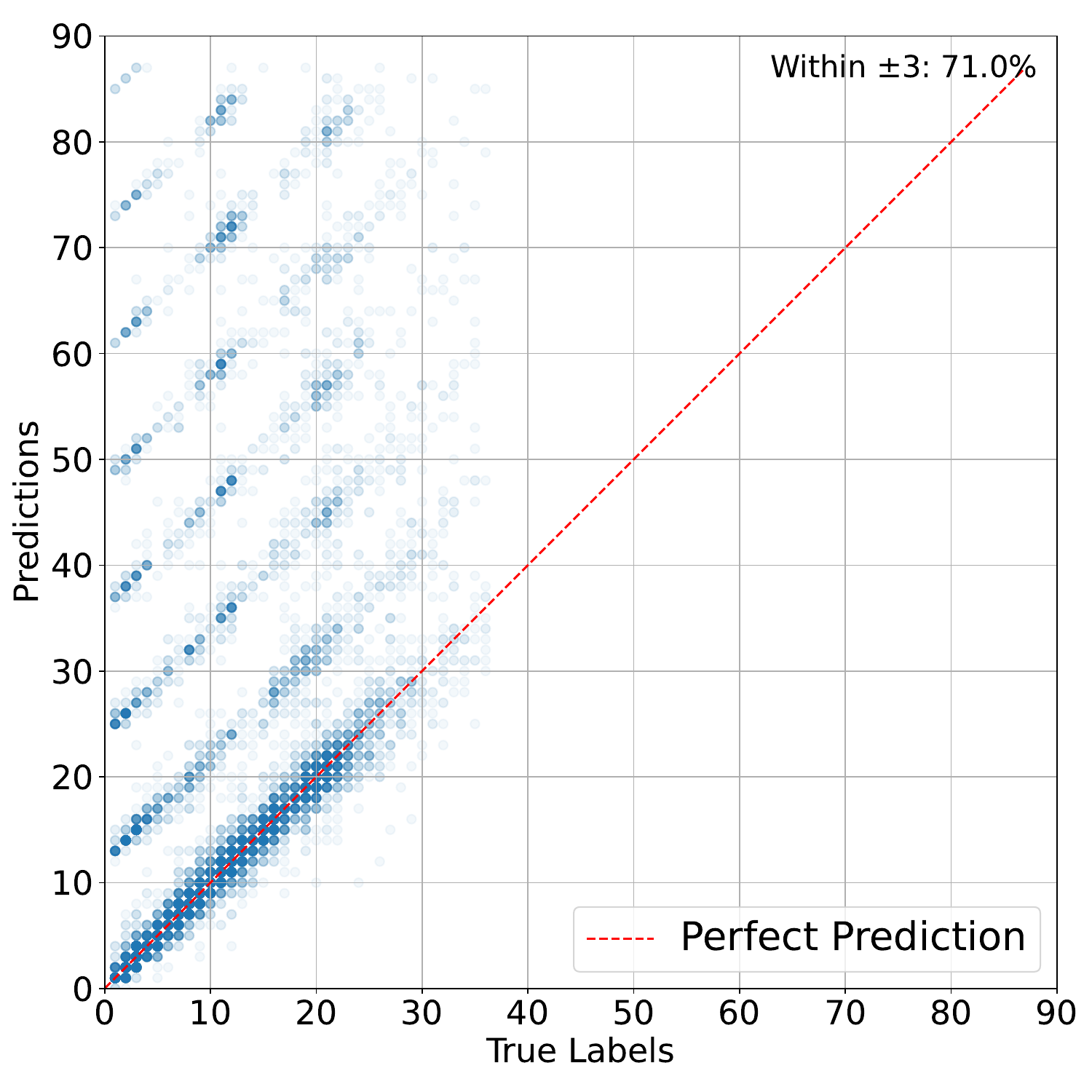}
    }
    \caption{Scatter plots showing prediction results: each point compares a trace’s predicted vs. true response count. Transparency ($0.05$) highlights density in overlapping regions.}
    \label{fig:traces-eval-images-03-model-03and01}
    \negspace
    \negspace
\end{figure}

\textbf{Per-Trace Prediction Accuracy.}  
Fig.~\ref{fig:traces-eval-images-03-model-03and01} presents scatter plots of predicted vs. true total responses for $T=0.1$\ seconds and $T=0.3$\,s windows. Each point represents one trace, and transparency ($\theta=0.05$) reveals areas of high point density. For example, if a trace is composed of five non-overlapping images with labels $1,0,2,4,1$ (total $8$) and model predictions $1,0,3,4,1$ (total $9$), it appears as $(8,9)$. Multiple traces with the same totals stack, increasing point opacity. 

For $T=0.1$\ seconds windows, the test set has 12,520 traces (avg. 21.2 images/trace); for $T=0.3$\,s windows, 12,142 traces (avg. 7.5 images/trace) are used. We use a $\pm3$ tolerance level because for both window lengths, the points represent the aggregated prediction sum and, thus, the aggregated errors as well. The $T=0.1$\ seconds model achieves 92.6\% accuracy versus 71\% for $T=0.3$\,s. Additionally, the $T=0.1$\ seconds predictions cluster more tightly along the diagonal, suggesting finer temporal granularity aids cumulative accuracy. This discrepancy arises from class imbalance and cumulative error accumulation in longer time windows.

\section{Conclusion}\label{conclusion}
This paper introduced VisQUIC, a large-scale dataset for encrypted QUIC traffic analysis, comprising 100,000+ labeled traces from over 44,000 websites. With encrypted traffic and SSL keys, VisQUIC enables in-depth studies of QUIC and HTTP/3 communications, providing a unique opportunity for granular encrypted traffic analysis.

A key contribution of VisQUIC is its integration of SSL keys and detailed metadata, enabling researchers to analyze encrypted traffic with controlled decryption and allowing them to rely solely on encrypted data. VisQUIC also introduces a novel image-based representation, transforming QUIC traffic into structured visuals. This enables ML models to analyze encrypted traffic, as shown by our benchmark algorithm, which achieved 97\% accuracy in HTTP/3 response estimation.

VisQUIC paves the way and provides a foundation for many promising research directions. Future work can develop privacy-preserving traffic analysis methods that balance security with analytical accuracy. It also allows studying QUIC’s behavior under different network conditions, browser implementations, and real-world use cases. Expanding VisQUIC to mobile and IoT traffic would enhance its applicability, providing deeper insights into QUIC’s role in today's networks. Future benchmarks can cover more challenges in encrypted traffic analysis, fostering the development of more sophisticated evaluation frameworks. By publicly releasing VisQUIC with full documentation and evaluation tools, we aim to accelerate encrypted traffic research and advance secure network protocols.

\appendix
\section{Appendices}\label{APP}
This section presents a custom loss function designed for benchmarking. The VisQUIC dataset presents a challenging benchmarking task for estimating the number of HTTP/3 responses in encrypted QUIC traffic. Traditional loss functions, such as cross-entropy or mean squared error (MSE), are inadequate for this task due to two key challenges: (1) class imbalance—where lower response counts dominate the dataset, leading to biased predictions—and (2) the ordinal nature of response counts, where the cost of misclassification depends on the numerical difference between predicted and actual values.

To address these challenges, we introduce a composite loss function integrating three components: a \textbf{Focused Loss (FL)} for class imbalance mitigation, a \textbf{Distance-Based Loss (DBL)} to penalize large deviations, and an \textbf{Ordinal Regression Loss (ORL)} to preserve response counts ranking relationships. The overall loss function is defined as:
\small
\begin{equation}
    L = \alpha \, \mathrm{FL}  + (1 - \alpha) \left(\beta \, \mathrm{ORL} + (1 - \beta) \mathrm{DBL}\right),
\end{equation}
\normalsize
where \( \alpha \) controls the balance between class weighting and ordinal constraints, and \( \beta \) determines the relative importance of ordinal ranking enforcement.

\textbf{Focused Loss (FL).}  
To address the heavy-tailed class distribution in HTTP/3 responses, we build upon focal loss~\cite{lin2017focal} by introducing a scaling factor that down-weights easy-to-classify samples. This ensures that harder-to-predict response classes receive greater attention during training:
\small
\begin{equation}
\mathrm{FL}(\mathbf{x}, \mathbf{y}) = \mathbb{E}_{(\mathbf{x}, \mathbf{y})} \left[ -w(y) \cdot \left(1 - \hat{\mathbf{y}}_{y}(\mathbf{x})\right)^\gamma \cdot \mathbf{y}^{\mathrm{T}}\log \hat{\mathbf{y}}(\mathbf{x}) \right],
\end{equation}
\normalsize
where \( w(y) \) is an inverse frequency weight for class imbalance, and \( \gamma \) controls the emphasis on hard-to-classify samples.

\textbf{Distance-Based Loss (DBL).}  
Since response counts are ordinal, the cost of misclassification should increase proportionally to the deviation from the ground truth. To incorporate this structure, DBL explicitly penalizes errors based on their absolute difference from the correct response count:
\small
\begin{equation}
\mathrm{DBL} = \mathbb{E}_{(\mathbf{x}, y)} \left[ 
\sum_{i} \hat{y}_i(\mathbf{x}) \cdot |i-y|
\right].
\end{equation}
\normalsize
This formulation ensures that small mispredictions receive lower penalties than large deviations, aligning model training with real-world tolerances in response estimation.

\textbf{Ordinal Regression Loss (ORL).}  
To reinforce ordinal constraints, we reformulate response estimation as a sequence of binary classification tasks, ensuring that predicted rankings maintain a consistent ordering:
\small
\begin{equation}
\mathrm{ORL} = \mathbb{E}_{(\mathbf{x}, \mathbf{y})} \left[ -\mathbf{y}^{\mathrm{T}}\log \sigma(\hat{\mathbf{y}}) - (1 - \mathbf{y})^{\mathrm{T}}\log \sigma(-\hat{\mathbf{y}}) \right],
\end{equation}
\normalsize
where \( \sigma \) is the sigmoid activation function. Unlike DBL, which penalizes based on numerical distance, ORL enforces ranking constraints to ensure predictions respect the ordinal structure of response counts.

The parameters \( \alpha \), \( \beta \), and \( \gamma \) control the relative influence of these components. Higher values of \( \alpha \) prioritize class balancing through FL, while lower values shift the emphasis toward ordinal consistency via DBL and ORL. The parameter \( \gamma \) adjusts the prioritization of difficult examples, making it particularly useful in highly imbalanced distributions.

This composite loss function helps VisQUIC-trained models optimize for accuracy while respecting both the ordinal nature of response counts and class imbalance. By integrating these components, the benchmark provides a standardized and robust evaluation framework for encrypted traffic analysis.


{\footnotesize
\bibliographystyle{IEEEtran}
\bibliography{main}
}
\end{document}